\documentclass[11pt,twoside]{article}

\usepackage{asp2004}
\usepackage{graphicx}
\usepackage{lscape}

\begin{document}

\title{SDSS~J0926+3624, the first eclipsing AM~CVn star, as seen with ULTRACAM}   %%% Fill in title

\author{T.R. Marsh$^1$, V.S. Dhillon$^2$, S.Littlefair$^2$, P.Groot$^3$,
P.Hakala$^4$, G.Nelemans$^3$,
G.Ramsay$^5$, G.Roelofs$^3$, D.Steeghs$^6$}   %%% Fill in author names

\affil{
$^1$Department of Physics, University of Warwick, Gibbet Hill Rd, Coventry
  CV4~7AL, UK. \\
$^2$Department of Physics, University of Sheffield, Sheffield
  S3~7RH, UK.\\
$^3$Department of Astrophysics, Radboud University Nijmegen, Toernooiveld~1,
NL-6525~ED Nijmegen, The Netherlands.\\
$^4$Tuorla Observatory, University of Turku,
Vaisalantie 20, FIN-21500 Piikkio,
Finland.\\ 
$^5$Mullard Space Science Laboratory/UCL, Holmbury St Mary, Dorking,
Surrey, UK.\\ 
$^6$Harvard-Smithsonian Center for Astrophysics, MS-57, 60~Garden Street, 
MA~02138, USA}    %%% Fill in author affiliations

\begin{abstract} %%% Abstract to run on from here.
We present light curves of SDSS~J0926+3624, the first eclipsing AM~CVn
star, observed with the high-speed CCD camera ULTRACAM on the WHT. We
find unusually that the accreting white dwarf is only partially
eclipsed by its companion. Apart from this, the system shows the
classic eclipse morphology displayed by eclipsing dwarf novae, namely
the eclipse of a white dwarf and accretion disc followed by that of
the bright spot where the mass transfer stream hits the disc. We are
able to fit this well to find masses of the accretor and donor to be
$M_1 = 0.84 \pm 0.05 \, \mathrm{M}_\odot$ and $M_2 = 0.029 \pm 0.02 \, \mathrm{M}_\odot$
respectively. The mass of the donor is significantly above its zero
temperature value and it must possess significant thermal content.
\end{abstract}

%%% MAIN BODY OF TEXT GOES HERE. CONSULT "INSTRUCTIONS FOR AUTHORS USING
%%% LATEX2E MARKUP", SECTIONS 2.3-2.6 FOR HELP WITH EQUATIONS, FIGURES,
%%% AND TABLES.

\section{Introduction}   %%% Top level section head (remove "%" symbol)
The AM~CVn stars are hydrogen-deficient interacting binaries. They have orbital
periods which range from 10 to 65 minutes, only possible because the absence of
hydrogen in their donor stars allows them to reach the high densities required at
short orbital periods. The accreting objects in these stars are white dwarfs, and they are
sometimes called ``double degenerates'' as the high densities of their mass donors
implies degeneracy pressure is important. It would be a mistake however simply to regard
these stars as two white dwarfs, as has sometimes happened in the past, because
it is quite possible for the donor stars to have significant thermal content,
and indeed the extent of this is one of the major questions concerning these
objects \citep{Deloye:entropy}. For a recent review of AM~CVn stars, see
\cite{Nelemans:AMCVnRev}.

The origin of the AM~CVn stars is uncertain. Three possible paths are under
debate. First is the merger of two white dwarfs under gravitational radiation
\citep{Nelemans:AMCVn}. This requires a mass ratio well below one for stability
\citep{Nelemans:AMCVn,Marsh:mdot}. The main uncertainty about this route is its
rate \citep{Marsh:mdot}. Second, is mass transfer from a core helium burning
star, e.g. an sdB star \citep{Iben:hestarcvs} although some degree of fine
tuning seems necessary to terminate the helium burning early enough that there
is no significant build-up of carbon. Third are cataclysmic variables with donor
stars that have left the main-sequence and built up a significant helium core
\citep{Podsiadlowski:AMCVn}. Here the problem seems to be that one would often
expect to see hydrogen. The second two paths reach a minimum period of about 10
minutes, whereas double white dwarfs merge at periods of 1 or 2 minutes, a
difference which is particularly significant in the context of the gravitational wave
observatory \emph{LISA} \citep{Nelemans:GWR}.

The donor stars in semi-detached binaries have a fixed density at any given
period. Completely degenerate, zero temperature stars obey a specific
mass-radius relation, and thus they can only fill their Roche lobe for a unique
mass at any one period. A star of finite entropy is always less dense than a
zero temperature star of the same mass, and therefore for a given orbital
period, finite entropy donors must be more massive than their degenerate
counterparts. The completely degenerate case therefore sets a lower limit to the
mass of the donor.  Since mass transfer in AM~CVn stars is thought to be driven
by gravitational radiation, then completely degenerate donors give the lowest
mass, least gravitational radiation, and lowest mass transfer rates. All this
shows the importance of mass measurements in these systems.  We have made
efforts to measure masses for these reasons \citep{Roelofs:AMCVn}, but the
measurements are reliant on several assumptions which are hard to verify.
Eclipsing systems always offer the most direct way to measure masses,
unfortunately, until recently not a single eclipsing AM~CVn had been discovered.

This changed with the discovery of SDSS~J0926+3624 by \cite{Anderson:AMCVn}.
SDSS~J0926+3624, discovered through its colour and spectrum, turned out to be an
eclipsing system with a period of $28\,$min. We therefore acquired photometry of
this object with the aim of carrying out an analysis similar to that applied to
hydrogen-dominated cataclysmic variables in the 1980s \citep[e.g.][]{Wood:ZCha}.

\section{Observations}
We observed SDSS~J0926+3624 on the nights of 1 to 3 March 2006 using
the high-speed CCD camera ULTRACAM mounted on the 4.2m WHT in La
Palma. ULTRACAM is a triple-beam camera and we observed through the
filters $u'$, $g'$ and $r'$. On each night we attempted to stay on the
target for long, uninterrupted runs to monitor both the eclipses and
the out-of-eclipse modulations. Table~\ref{tab:log} shows a log of the
observations.
\begin{table}[t]
\centering
\begin{tabular}{lccl}
\hline
Date          & \multicolumn{2}{c}{Time}    & Comment \\
start         &  start & end                & \\
\hline
01 Mar 2006   & 22:28  & 04:48              & Poor weather at start \\
02 Mar 2006   & 20:04  & 04:49              & Seeing $\sim 1$ --
$1.5$'', clear\\
03 Mar 2006   & 19:57  & 03:59              & Variable seeing, $1.2$
-- $2.0$''\\
\hline
\end{tabular}
\caption{Log of the observations. \label{tab:log}}
\end{table}

We used an exposure time of $3\,$sec for the majority of our
observations; dead time was $0.024\,$sec for our frame transfer CCDs.
The data were reduced with standard aperture photometry. Observations
of a standard star showed that SDSS~J0926+3624 had $g' \approx 19.3$
out of eclipse, as observed by \citet{Anderson:AMCVn}.

\section{Results}
\begin{figure}[tb]
\centering
\includegraphics[width=0.98\textwidth]{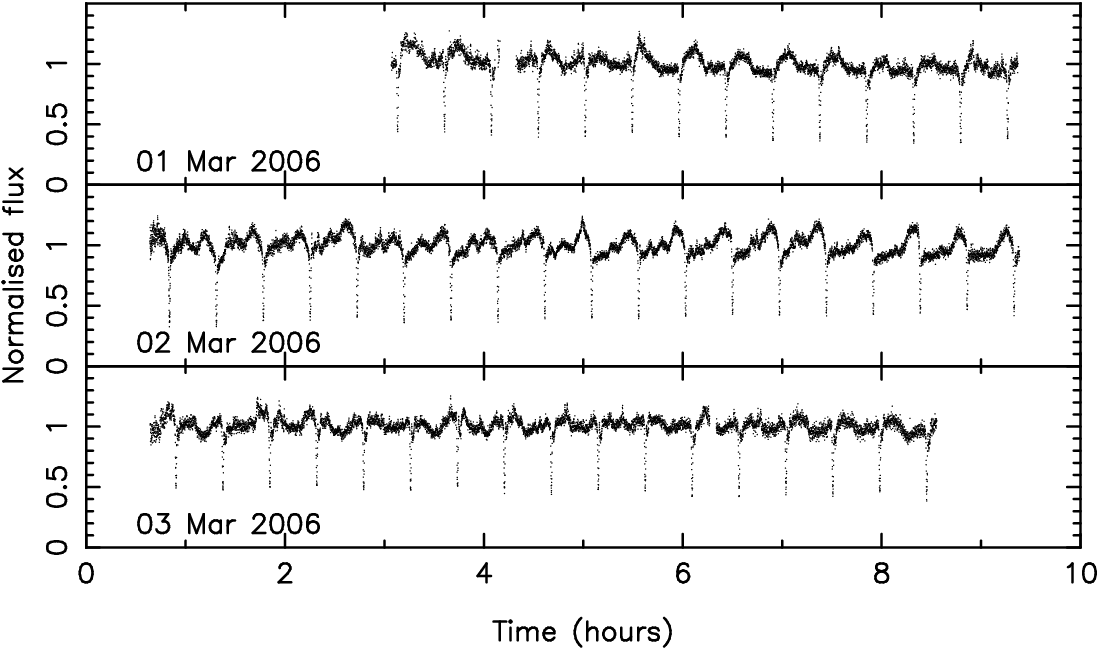}
\caption{The light-curves of SDSS~J0926+3624 ($u'$, $g'$ and $r'$
  combined) over the three nights of our run. The mean level outside
  eclipse has been normalised to one. \label{fig:all}}
\end{figure}

Fig.~\ref{fig:all} 
shows all the data from the run, and covers 50 eclipses of
SDSS~J0926+3624. The eclipse depth is variable, but is as much as 70\%
of the out-of-eclipse flux. There is marked variability outside
eclipse which changes character from night to night. This is caused by
what are known as ``superhumps'', periodic flaring of the accretion on a
period slightly longer than the orbital period, and thought to be
caused by a tidally-driven instability of discs in extreme mass ratio
systems \citep{Whitehurst:superhumps}. 

\begin{figure}[tb]
\centering
\includegraphics[angle=270,width=0.98\textwidth]{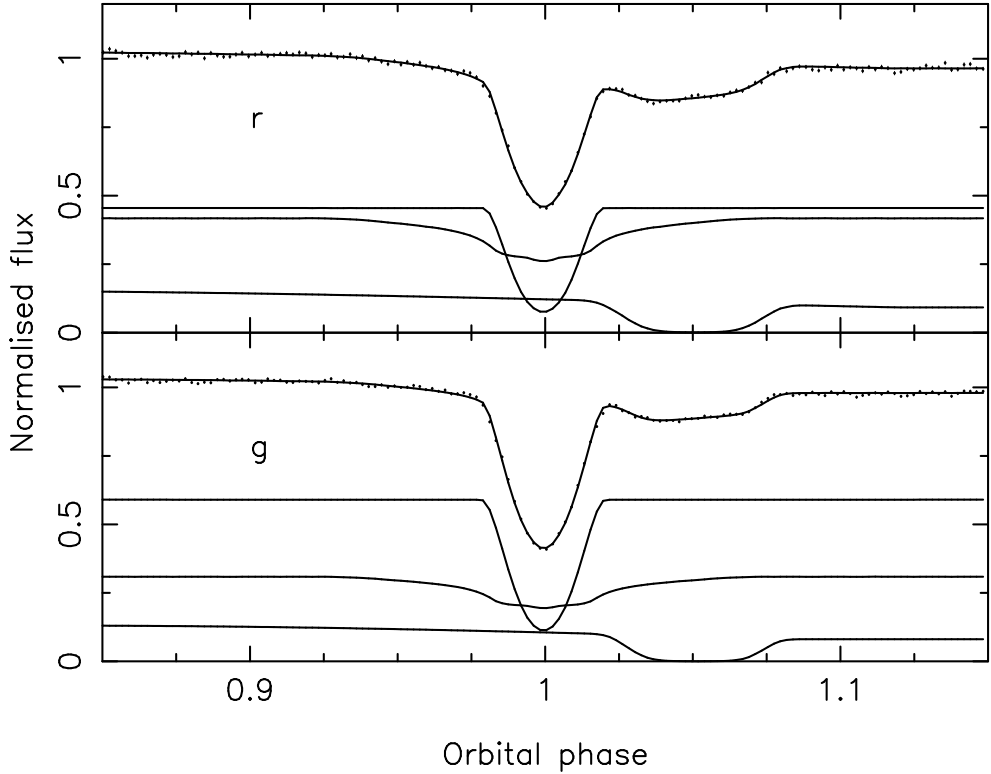}
\caption{The phase-folded light-curves of SDSS~J0926+3624 in $g'$ and $r'$
based upon the first and last nights' data. Solid lines show
calculations
of the eclipse of the white dwarf (narrowest), the bright-spot (offset
from phase zero) and the accretion disc (shallowest) as well as the
sum of all three.
\label{fig:fold}}
\end{figure}

We removed most of the effect of the superhump by fitting and
subtracting a sinusoid to the data outside eclipse. We then folded the
first and last nights' data, which have a similar morphology, on the
orbital period to obtain the data plotted in Fig.~\ref{fig:fold}.
The data show a relatively narrow eclipse, lasting approximately from
phase $-0.02$ to $+0.02$, or a little over one minute in time. It is
no coincidence that it is centred on phase zero as this feature was
used to define the conjunction phase as we believe it is the white
dwarf. A remarkable feature is that it is round-bottomed, indicating
that we are seeing the \emph{partial} eclipse of the accretor in this 
system.

Following the narrow eclipse, there is another, broader feature. This
is the eclipse of the bright spot where the mass transfer stream hits
the disc. Its extreme displacement from the white dwarf eclipse such
that its ingress starts after the white dwarf's egress is a sign that
the system is of extreme mass ratio because in such systems, the high
angular momentum of the transferred gas deflects it far in advance of
the mass donor's orbit as it moves towards the accretor. The bright
spot is eclipsed totally, but its eclipse is not flat-bottomed because of the
eclipse of a third component, the accretion disc. This is coming out
of eclipse during the eclipse of the bright spot, and hence the
increasing flux at this time.

\begin{figure}[tb]
\centering
\includegraphics[angle=270,width=0.98\textwidth]{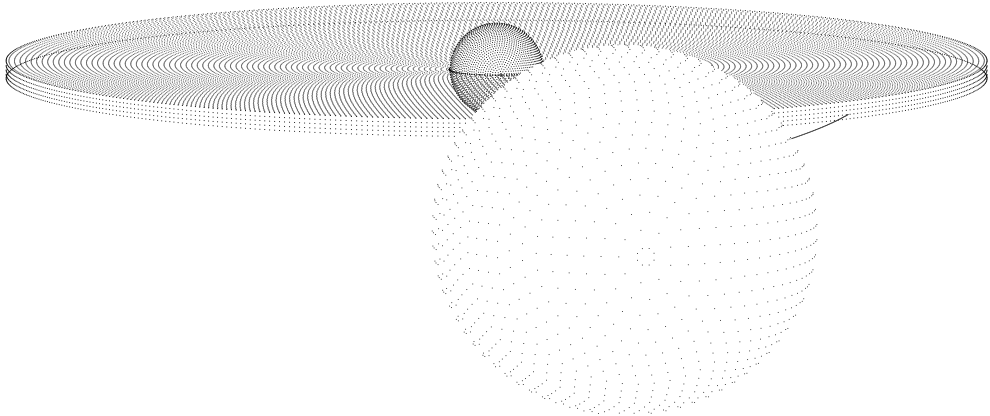}
\caption{A visualisation of the relative sizes of the components used to
model the light-curves of Fig.~\protect\ref{fig:fold}. The line leading from the
right of the donor star (closest to us) is the mass transfer stream showing that
the spot where it hits the disc has yet to be eclipsed at this phase when the
accretor has almost emerged from eclipse. Note that the accretor and donor stars
are not that different in size, which is a feature of these very compact binary stars.
\label{fig:geom}}
\end{figure}

We modelled all these features with a light-curve fitting program
which sets element grids over all components and computes their mutual
occultations (Fig.~\ref{fig:geom}), 
accounting for the Roche-distorted geometry. The parameters were then
adjusted to obtain the best fit using Levenberg-Marquart minimisation
of $\chi^2$. The results from this are plotted in
Fig.~\ref{fig:fold}, confirming the partial eclipse of the white
dwarf and total eclipse of the bright spot. The fit was obtained for a
mass ratio $q = M_2/M_1 = 0.035 \pm 0.002$, an inclination $i =
83.1\pm0.1^\circ$ and a relative radius of the accretor of $R_1/a =
0.033 \pm 0.002$. Using a zero-temperature relation for the
mass-radius relation of the donor (which will be improved upon in the
future) leads to $M_1 = 0.84 \pm 0.05 \, \mathrm{M}_\odot$ and $M_2 =
0.029 \pm 0.02 \, \mathrm{M}_\odot$ for the accretor and donor
respectively. The uncertainties are preliminary, nonetheless we believe that
these are the most secure masses for any AM~CVn star. We note that accounting
for  the finite temperature of the accretor will lead to a small increase in 
the masses.

The mass of the donor is of particular interest. Had it been completely
degenerate, it should have had a mass close to $0.020\, \mathrm{M}_\odot$. 
Instead it is substantially higher than this indicating a significant level of thermal 
energy \citep{Deloye:entropy}. The implications of this for the evolution of the 
binary need to be worked out.

An interesting point from our model of the light-curve is that the white
dwarf contributes  most of the optical flux in this system. This is comparable
to quiescent dwarf novae.  However, unlike the great majority of quiescent dwarf
novae, SDSS~J0926+3624 shows obvious superhumps in this state. Superhumps in
dwarf novae are normally associated with outburst bright states. A likely
explanation for this lies in the extreme mass ratio of the system, although
there have not been any reports as far as we are aware of superhumps in similar
AM~CVn stars in their low state.

\begin{figure}[tb]
\centering
\includegraphics[angle=270,width=0.98\textwidth]{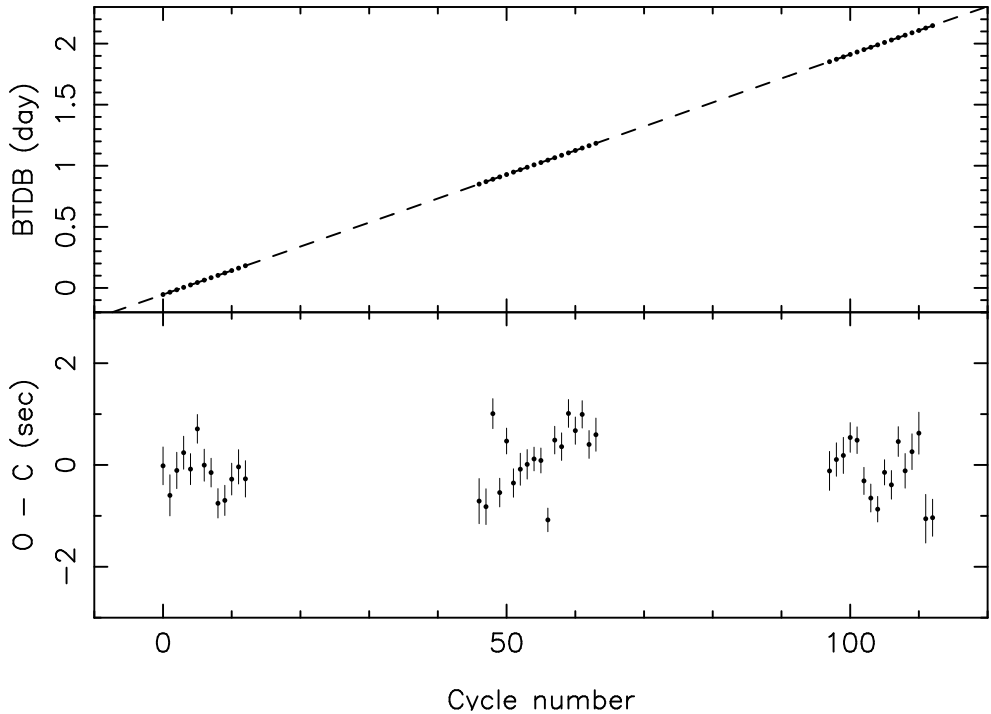}
\caption{Eclipse times based upon our model of the light curve. Systematics in
  the residuals in the middle night are likely the result of the superhumps.
\label{fig:times}}
\end{figure}

We finish by showing the times of the eclipses in Fig.~\ref{fig:times}.
The mean eclipse time has an uncertainty of $\approx 0.2\,$sec, which shows that
it should be possible to detect period changes of plausible magnitude.
Gravitational radiation, for example, should cause $\sim 5\,$sec departure from
linearity over 10 years.

\section{Conclusions}
We have presented high-speed light curves of the first eclipsing AM~CVn star,
SDSS~J0926+3624 which we find to display the usual form of eclipse light curve
displayed by eclipsing dwarf novae. Modelling of the light curve leads to an
extreme mass ratio of $q = 0.035 \pm 0.002$, as expected for these systems and
a donor mass 50\% above the mass expected for pure degeneracy indicating that
thermal pressure is significant. 

\acknowledgements 
TRM was supported by a PPARC Senior Research Fellowship during the
course of this work, ULTRACAM was made possible by a PPARC grant
and SL was supported by a PPARC PDRA.

%%% THE BIBLIOGRAPHY
%%%
%%% CONSULT SECTION 3 OF "INSTRUCTIONS FOR AUTHORS" FOR HOW TO USE NATBIB.
%%% AUTHORS ARE ENCOURAGED TO USE EITHER THE "THEBIBLIOGRAPY" ENVIRONMENT
%%% BY UNCOMMENTING (DELETING THE "%" SYMBOL) THE COMMANDS BELOW, OR BY
%%% USING THE BIBTEX ENVIRONMENT. TO FIND OUT WHICH IS APPLICABLE TO YOUR
%%% CONTRIBUTION, CONSULT THE VOLUME EDITORS FOR YOUR PROCEEDINGS.
%%%

%\bibliographystyle{mybst}

%\bibliography{refs}

%\begin{thebibliography}{}
%\bibitem[]{}
%\bibitem[]{}
%\bibitem[]{}
%\bibitem[]{}
%\bibitem[]{}
%\bibitem[]{}
%\bibitem[]{}
%\bibitem[]{}
%\bibitem[]{}
%\bibitem[]{}
%\bibitem[]{}
%\bibitem[]{}
%\end{thebibliography}

\end{document}